\documentstyle[floats,aps,pre,psfig]{revtex}
\newcommand{\be}{\begin{equation}}
\newcommand{\ee}{\end{equation}}
\newcommand{\bea}{\begin{eqnarray}}
\newcommand{\eea}{\end{eqnarray}}
\newcommand{\les}{\ell_{\hbox{\tiny ES}}}
\newcommand{\ld}{\ell_{\hbox{\tiny D}}}
\newcommand{\lnuc}{\ell_{\hbox{\tiny nuc}}}
\newcommand{\lnt}{\ell_{\hbox{\tiny nuc}}^{\hbox{\tiny T}}}
\newcommand{\lnv}{\ell_{\hbox{\tiny nuc}}^{\hbox{\tiny V}}}
\newcommand{\lnb}{\ell_{\hbox{\tiny nuc}}^{\hbox{\tiny B}}}
\newcommand{\pnuc}{p_{\hbox{\tiny nuc}}}
\newcommand{\peffn}{p_n^{\hbox{\tiny eff}}}
\newcommand{\peff}{p^{\hbox{\tiny eff}}}
\newcommand{\peffx}{p^{\hbox{\tiny eff}}(x)}
\newcommand{\ps}{p^{\hbox{\tiny S}}}
\newcommand{\pu}{p^{\hbox{\tiny U}}}
\newcommand{\Pdep}{P_{\hbox{\tiny dep}}}
\newcommand{\Pn}{P^{\hbox{\tiny (N)}}}
\newcommand{\pone}{p^{\hbox{\tiny (1)}}_n}
\newcommand{\ptwo}{p^{\hbox{\tiny (2)}}_n}

\newcommand{\oni}{\omega_{\hbox{\tiny NI}}}
\newcommand{\omf}{\omega_{\hbox{\tiny MF}}}
\newcommand{\ore}{\omega_{\hbox{\tiny RE}}}
\newcommand{\ttr}{\tau_{tr}}
\newcommand{\tdep}{\tau_{dep}}
\newcommand{\tres}{\tau_{res}}
\newcommand{\Nall}{N_{\hbox{\tiny all}}}
\newcommand{\Ndis}{N_{\hbox{\tiny dis}}}
\newcommand{\Wni}{W_{\hbox{\tiny NI}}}
\newcommand{\Pni}{P_{\hbox{\tiny NI}}}
\newcommand{\PNI}{P^{\hbox{\tiny NI}}}

\newcommand{\fra}[2]{\hbox{${#1\over #2}$}}
\newcommand{\bold}[1]{\hbox{\bf x}}

\begin{document}
\draft

\title{The process of irreversible nucleation in multilayer growth.\\
I. Failure of the mean-field approach.}

\author{Paolo Politi$^{1,\dag}$ and Claudio Castellano$^{2,*}$}
\address{$^1$ Istituto Nazionale per la Fisica della Materia, Unit\`a di
Firenze, Via G. Sansone 1, 50019 Sesto Fiorentino, Italy}
\address{$^2$ Istituto Nazionale per la Fisica della Materia, 
Unit\`a di Roma 1\\ and
Dipartimento di Fisica, Universit\`a di Roma
``La Sapienza'', Piazzale Aldo Moro 2, 00185 Roma, Italy}

\author{(\today)}
\author{\parbox{397pt}{\vglue 0.3cm \small
The formation of stable dimers on top of terraces during epitaxial
growth is investigated in detail.
In this paper we focus on mean-field theory, the standard approach
to study nucleation. Such theory is shown to be unsuitable
for the present problem, because it
is equivalent to considering adatoms as independent diffusing particles.
This leads to an overestimate of the correct nucleation
rate by a factor ${\cal N}$, which has a direct physical meaning:
in average, a visited lattice site is visited ${\cal N}$ times
by a diffusing adatom.
The dependence of ${\cal N}$
on the size of the terrace and on the strength of step-edge barriers
is derived from well known results for random walks.
The spatial distribution of nucleation events is shown
to be different from the mean-field prediction, for the same
physical reason.
In the following paper we develop an exact treatment of the problem.
}}
\maketitle

\section{Introduction}
\label{Intro}

A crystal can be produced artificially with different
techniques: we can pull it from the melt,
grow it from a solution or obtain it via deposition from a gas/vapor
phase onto a suitable substrate.
One of the key mechanisms of the growth process is the formation of 
supercritical nuclei, that is the nucleation -- via diffusion and
aggregation -- of crystalline clusters whose growth rate exceeds the
decay rate.

In this paper we  devote our attention to epitaxial growth
by atomic or molecular beams~\cite{libroMBE}: 
particles travel ballistically towards
the growing surface where they undergo a thermally activated diffusion
process. The size $i^*$ of the critical nucleus is typically a few units
and its actual value depends on several factors: 
the substrate and the type of adatoms deposited
determine the activation barriers for the different atomistic
processes, while the temperature and the flux determine what processes
are really relevant on the time scale of the experiment.
Here we consider the simplest case: 
nucleation is irreversible ($i^*=1$), i. e. once
two adatoms meet they form a stable dimer.
After nucleation, the stable nucleus grows by capturing other adatoms. 

We start by explaining in general terms the role of nucleation in the
different stages of epitaxial growth and by discussing
how rate equations and mean-field theory deal with it.
The focus of the rest of the paper will be on nucleation on
top of terraces bound by descending steps 
(often called second layer nucleation).

In the submonolayer regime diffusion takes place on the substrate:
adatoms are deposited randomly and they diffuse until they meet
another wandering adatom or a growing cluster. 

Rate equations~\cite{rate_eq} are widely used to describe the processes 
of adatom capture and adatom detachment from clusters of $j$ atoms
($j-$clusters, $j\ge 2$).
If only adatoms are mobile, the spatial density $\rho_j$ of
$j-$clusters varies in time according to the relation~\cite{libro_Venables}
$\partial_t \rho_j = U_{j-1} - U_j$ where $U_j$ is 
the net rate of the process ($j-$cluster)$\Rightarrow$($j+1$-cluster).
$U_j$ is the sum of a ``growth term" and a ``decay term":
the growth term represents the aggregation of an atom into a 
$j-$cluster and it has the form $\sigma_j D\rho \rho_j$, 
where $\rho$ is the adatom density, $D$ the adatom diffusion constant
and $\sigma_j$ is an adimensional capture number;
the decay term represents the thermal detachment of an atom from a 
$(j+1)-$cluster and it has the form $-\left(\fra{4D}{a_0^2}\right)
\rho_{j+1}\exp(-\Delta E/T)$, where $a_0$ is the lattice constant and
$\Delta E$ is the energy difference between a $j-$cluster 
(plus a free adatom) and a $j+1$-cluster.
If $i^*=1$ such term is absent, because all $j-$clusters are stable
and the nucleation rate is 
$\ore =\sigma_1 D \rho^2$ (the subscript standing for ``Rate Equations").

The capture factor $\sigma_j$ is defined through the flux $\Phi_j$ of
adatoms attaching to the $j-$cluster: $\Phi_j\equiv D\sigma_j \rho$,
and it accounts for the different adatom densities surrounding
islands of different size. In a Mean-Field (MF) approach such density 
is taken as a constant, $\sigma=2d$~\cite{nota_sigma} 
and the nucleation rate {\it per}
lattice site reads $\omf = 2dD \bar\rho^2$.

By monitoring the adatom density $\bar\rho$ and the total density of islands
$\bar\rho_{\hbox{\tiny TOT}} = \sum_{j\ge 2} \bar\rho_j$ one
realizes the existence of a regime characterized by
$\bar\rho_{\hbox{\tiny TOT}}$ almost constant in time, prior to the 
coalescence regime~\cite{AFL}.
The average distance $\ld$ between islands can therefore be defined:
$\ld^d = 1/\bar\rho_{\hbox{\tiny TOT}}$,
$d$ being the dimensionality of the substrate: usually $d=2$,
but in the following we will consider $d=1$ as well.
$\ld$ is called the diffusion length and it also gives the typical 
linear distance travelled by an adatom before being incorporated in an island.

During the time $1/F$, necessary for the deposition of a monolayer, there is
on average a nucleation event per area $\ld^d$~\cite{VPW}. Therefore, 
according to mean-field theory, 
$\omf\cdot\ld^d/F\sim 1$, i. e.
$D\bar\rho^2 \ld^d /F\sim 1$ with (see App.~\ref{App_rho})
$\bar\rho\sim\fra{F}{D}\ld^2$. So we obtain the result
$\ld \sim (D/F)^\gamma$, with $\gamma=\fra{1}{5}$ in $d=1$
and $\gamma=\fra{1}{6}$ in $d=2$.
This relation is wrong in one dimension,
but essentially correct in two dimensions~\cite{VPW,PVW}.
In Sec.~\ref{nuc_rate} it will become clear that Mean-Field Theory (MFT)
gives the correct result for $\gamma$ in $d=2$ because step-edge
barriers at island edges play no role in the submonolayer regime.

If adatoms freshly attached to a growing nucleus are immobile
the resulting island is normally fractal, but
if they can diffuse along the step-edge the island gets compact.
Compact islands are therefore obtained at not
too low temperatures (so that thermally activated edge diffusion
actually occurs) whilst nucleation is irreversible at not
too high temperatures (so that a dimer is thermally stable): 
in some experimental systems both
conditions are fulfilled in a given temperature range:
Pt(100)/Pt in the range 350-430 K~\cite{Pt}, Ag(100)/Ag
in the interval 200-300 K~\cite{Ag}, and Fe(100)/Fe
between 300 and 500 K~\cite{Fe} are some examples.

Once island coalescence has set off, most nucleation events
take place on top terraces. The obvious reason is that even for 
layer-by-layer growth a new atomic layer starts to 
form before the completion of the previous one
and consequently the growing surface is made up of a certain
number of exposed layers.
Terraces can be classified as top (T), vicinal (V) and bottom (B) terraces
according to the type of steps surrounding them: in general only top
terraces attain a size large enough to have a considerable probability
of nucleating a new island. Since this probability grows abruptly
from zero to one with increasing terrace size~\cite{TDT},
it is possible to introduce a critical nucleation length $\lnuc$~\cite{EV}.
In the presence of step-edge barriers~\cite{seb}, 
hindering interlayer transport,
such length differs for the three types of terraces:
$\lnt$ goes to zero with increasing barriers while
$\lnv$ and $\lnb$ remain finite in such limit~\cite{EV,review}. 
Nucleations generally occur on top terraces because
the size of vicinal and bottom terraces hardly reaches the
nucleation length. However nucleation is a stochastic phenomenon
and therefore dimer formation may also occur on vicinal
terraces every now and then, while on bottom terraces it is an exceedingly
rare event. These occasional ``vicinal nucleations" are an important
stabilizing mechanism of the surface, even if
their relative weight decreases as barriers increase~\cite{PV,nuc_vic}.

The mean-field approach to evaluate the nucleation rate for
(reversible and irreversible) nucleation on top terraces has been
worked out by Tersoff, van der Gon and Tromp in Ref.~\cite{TDT} and since
then their results have been widely used to analyze experimental data and
extract the values of step-edge barriers.
The mean-field approximation can also be used in `mesoscopic'
models of growth~\cite{Ratsch} for implementing the rule for
the spatial distribution $P(x)$ of nucleation events, via
the relation $P(x) \propto \rho^2(x)$.

Recently, several authors~\cite{Maass,KPM,Krug2nd} have started to review
critically the MF approach. 
In Refs.~\cite{Maass,Krug2nd} authors are mainly interested in the
validity of MFT for different values of the critical size $i^*$:
this check has been done through scaling analysis and kinetic
Monte Carlo simulations.
A deeper investigation of the irreversible case ($i^*=1$)
is done in Ref.~\cite{KPM} in the limit of strong step-edge barriers,
through a quantitative approach based on the different time scales involved
in the nucleation process.

The inaccuracy of MF theory in dealing with the problem of second layer
nucleation has therefore already been pointed out in the literature.
In this work we analyze thoroughly the validity of MF theory
for irreversible second layer nucleation, 
we rigorously prove its inadequacy both
for the nucleation rate $\omega$ and the spatial distribution $P(x)$
of nucleation sites
and understand the physical origin of its failure.
In the following paper~\cite{Politi01b} we compute exactly $\omega$ and $P(x)$,
providing the correct expressions that must be used in place of the
MF approximations.

We have organized this paper as follows.
The three relevant time scales for the nucleation process
and the associated regimes are introduced
and discussed in the next Section.
In Sec.~\ref{gen_frame} we explain how we can get rid of the stochastic
nature of the deposition process~(\ref{riduzione}) and we introduce
the quantities of interest in the paper, 
the spatial distribution of nucleation events (\ref{spaziotempo})
and the nucleation rate (\ref{ss_omega}).
In Sec.~\ref{mft} we show the equivalence of MF theory with a model of
noninteracting particles, both for the nucleation rate 
(\ref{nuc_rate}) and for the spatial distribution (\ref{spat_dist}). 
The final Section contains a critical summary of the results.

A short report of this work has been published in
Ref.~\cite{Castellano01}.

\section{Time scales}
\label{time_scales}

We consider a top terrace of fixed linear size $L$, subject to a flux
$F$ of particles.
Once on the terrace, each particle moves with a diffusion constant $D$
until it meets another particle or it leaves the terrace.
At step edges an additional energy barrier, usually called 
Ehrlich-Schwoebel (ES) barrier~\cite{seb},
reduces the rate $D'$ of interlayer transport and the ES length
\be
\les = \left( {D \over D'} -1 \right) a_0
\ee
measures the asymmetry between $D$ and $D'$.
In the following the lattice constant $a_0$ will be taken as
unit length ($a_0=1$).

In general, throughout the paper we will consider discrete space and time,
i. e. particles moving on a lattice (a square lattice in $d=2$),
at fixed time steps.
However we will  sometimes use a continuum notation as well.
The matching between discrete and continuum is
straightforward:
if $t_c$ is a continuous time and $t_d$ its discrete
counterpart (number of time steps), then $t_c=t_d\cdot\Delta t$, where 
$\Delta t =1/(2dD)$ is the time an atom remains on each lattice site.

Let us now discuss the three time scales involved in
the problem~\cite{KPM}.

\noindent $\bullet$
The traversal time ($\ttr$) is the average time needed by a diffusing
particle to reach the terrace boundaries. In the large $L$ limit
\be
\ttr = \beta L^2/D~ ,
\label{eq_ttr}
\ee
where $\beta$ is a numerical prefactor depending on the dimension
$d$ and the shape of the terrace.
Its value is discussed after Eq.~(\ref{eq_tres}).

\noindent $\bullet$
The residence time ($\tres$) is the average time a particle spends
on the terrace. It is related to the average density $\bar\rho$ 
of adatoms via the relation~\cite{KPM} $\bar\rho = F\tres$, where
the density $\rho$ (and therefore its average value $\bar\rho$) 
can be determined (see App.~\ref{App_rho}) 
by solving the stationary diffusion equation in the presence
of a constant flux $F$: 
\be
D\nabla^2\rho +F =0 ~.
\label{eq_dif}
\ee

Boundary conditions depend on the strength of the Ehrlich-Schwoebel (ES) 
barrier at descending steps.
If the equilibrium adatom density -- due to thermal detachment
from steps -- is neglected, the boundary condition for $\rho$
is: $\partial_\perp\rho|_{\hbox{\tiny step}} = 
\fra{\rho}{\les}|_{\hbox{\tiny step}}$, where $\partial_\perp$ is the local
derivative in the direction perpendicular to the step
(directed inward the terrace).
In a discrete picture (for example in $d=1$) if $n=1$ is a lattice site
at the edge of the terrace and $n=0$ is its fictitious neighbour
outside the terrace, we have $\rho(1)-\rho(0)=\rho(0)/\les$, i. e.
$\rho(0) = \fra{\les}{1+\les}\rho(1)
\equiv a\rho(1)$.

The solution of Eq.~(\ref{eq_dif}) (see App.~\ref{App_rho}) 
gives the following result, valid both in one and
two dimensions: $\bar\rho  = \fra{F}{D}( \beta L+\alpha\les) L$, where 
$\alpha$ is another numerical factor 
depending on the dimension $d$ and on the shape of the terrace.
We can finally write:
\be
\tres = (\beta L+\alpha\les) L/D ~.
\label{eq_tres}
\ee
In the absence of ES barriers ($\les=0$) $\tres$ and $\ttr$ are 
equal. From
Eqs.~(\ref{rho_1d},\ref{rho_2d},\ref{tresgeneric},\ref{A54}) 
we infer that
in $d=1$, $\beta=1/12$ and $\alpha=1/2$. In $d=2$,
for a circular terrace of radius $L$, $\beta=1/8$ and $\alpha=1/2$, 
while for a square terrace, $\beta \simeq 32/\pi^6$ and $\alpha=1/4$.

In the discrete picture $\tres$ is clearly equal to the total number
of sites ($\Nall$)
visited by an atom during its diffusional motion on the terrace. Since
the adatom stays on a lattice site a time $1/(2dD)$ 
we have $\tres=\Nall/(2dD)$.
Hence, the residence time $\tres$, the average density $\bar\rho$
and the number $\Nall$ of all sites visited by the adatom carry the
same piece of physical information, once $F$ and $D$ are set. 
The quantity $\Nall$ should not be confused with
the number of {\it distinct} sites ($\Ndis$) visited by an adatom:
a given lattice site, visited $k$ times, contributes for 1 to
$\Ndis$ and for $k$ to $\Nall$.

\noindent $\bullet$
The deposition time ($\tdep$) is the average time
between a deposition event and the next one.
For a terrace of area ${\cal A} = L^d$:
\be
\tdep = {1\over F {\cal A}} = {1 \over F L^d}~.
\label{eq_tdep}
\ee

Physically sensible values for $F$, $L$ and $D$ imply that $\ttr\ll\tdep$. 
This relation indeed is $\fra{L^2}{D}\ll\fra{1}{FL^d}$, i. e.
$\fra{D}{F}\gg L^{d+2}$: we can now recall 
the diffusion length~\cite{libroJV} introduced in Sec.~\ref{Intro},
$\ld\sim(\fra{D}{F})^\gamma$ and
measuring the `maximal' size of a terrace {\it in the absence} of step-edge
barriers.
For irreversible nucleation the exponent $\gamma$ is equal
to~\cite{libroJV} $\fra{1}{2(d+1)}$ so that we obtain the condition
$\ld^{\fra{2(d+1)}{d+2}}\gg L$, i. e. $\ld^{4/3}\gg L$ in $d=1$ and
$\ld^{3/2}\gg L$ in $d=2$. 
Smooth growth requires that $\ld\gg 1$; furthermore $L$ is at most of order
$\ld$ if $\les=0$, but for finite barriers it is (much) smaller.
We conclude that the above conditions are fulfilled and that we can
safely suppose that $\ttr\ll\tdep$.

This inequality has a consequence of primary
importance: processes involving more than two adatoms at a time can be 
fully neglected. Once two adatoms are simultaneously on the terrace, they meet
-- if they do -- on the time scale of the traversal time $\ttr$.
This fact is intuitively clear and it is proven in Ref.~\cite{Politi01b}.
The probability that a third atom lands in the meanwhile is $\ttr/\tdep$,
negligibly small. Irreversible nucleation is therefore the
result of two-adatoms processes only.

Depending on the relative size of $\tres$ with respect to the other two 
time scales, three different regimes may occur:
\bea
i) ~~~~~ 
& \ttr \simeq \tres \ll \tdep &
\hbox{~~~~Zero or weak barriers} \\
ii) ~~~~~ 
& \ttr \ll \tres \ll \tdep &
\hbox{~~~~Strong barriers} \\
iii) ~~~~~ 
& \ttr \ll \tdep \ll \tres &
\hbox{~~~~Infinite barriers}
\eea

The difference between the three regimes is easily understood.
A nucleation may occur only if a new adatom is deposited
before the previous one leaves the terrace. If $\tres \ll \tdep$
(regimes {\it i} and {\it ii}) this is a rare event.
When it happens, the second atom finds the
first one with a spatial distribution that differs in cases {\it i}
and {\it ii} (see App.~\ref{App_eff}).
If $\les\ll L$ (regime {\it i}) when the adatom reaches the edge of the
terrace, it gets off. Steps act as absorbing boundaries
and the adatom density vanishes there: $\rho$ has a parabolic shape
with a maximum in the middle of the terrace (see App.~\ref{App_rho}).
If $\les\gg L$ (regime {\it ii})
the adatom is pushed back from the terrace edge
several times before being able to descend. Steps act as (imperfect)
reflecting walls and $\rho$ is approximately uniform
over the whole terrace.  
In regime {\it iii} when a new adatom is deposited it always finds the 
previous one still on the terrace, they both have a flat distribution and 
they will certainly meet.

\section{General formalism}
\label{gen_frame}

In the present and in the following paper we are going to use 
a discrete formulation for particle dynamics, both in space and
in time. In this Section $n$ indicates the whole set of $d$
integer numbers specifying the position of a particle on the terrace.
A nucleation event is assumed to occur   
when two adatoms are on the same lattice site, rather than
on neighbouring lattice sites:
this definition avoids useless mathematical complications,
but retains all the physics of the nucleation process.

\subsection{Reduction to two particles deposited simultaneously}
\label{riduzione}

It is clear that the problem of 
dimer formation on a terrace involves the study of the diffusion of
two particles deposited {\em at different times}: the spatial and
temporal distributions of landing events play therefore a prominent role.
The incoming flux of particles is supposed to be spatially and
temporally uniform~\cite{nota_uniforme}: a particle arrives
on each lattice site with uniform probability $\pu_n \equiv 1/L^d$
and the interarrival time $\tau$ between two deposition events 
decays exponentially~\cite{nota_dep}:
\be
\Pdep(\tau) = \exp(-\tau/\tdep)/\tdep~.
\label{Pdep}
\ee

Let us now consider any ``two-particles" quantity ${\cal O}$,
i. e. any quantity depending on the initial distributions of particle
(1) and (2) and on their interarrival time $\tau$.

Let $p_n(0)=\pu_n=1/L^d$ be the initial uniform distribution of
an atom and $p_n(\tau)$ its dynamical evolution at time $\tau$
({\it in the absence} of other particles).
If particle (1) is deposited at time zero and particle (2) a
time $\tau$ later, we call ${\cal O}\{\pone(\tau),\ptwo(0)\}$
the resulting physical quantity.
${\cal O}$ might be, for example, the probability $\pnuc$ that a deposited
particle nucleates a dimer before getting off the terrace
[see Eq.~(\ref{pnucexpl})].
Once ${\cal O}\{\pone(\tau),\ptwo(0)\}$ is known, one should
evaluate its average over $\tau$:
\be
\hat {\cal O} = \sum_{\tau=0}^{\infty} \Pdep(\tau)
{\cal O}\{\pone(\tau),\ptwo(0)\}~.
\ee

The crucial point is that if ${\cal O}$ is {\it linear} in the initial
distributions $p^{\hbox{\tiny (1,2)}}_n$ of the two atoms
(as all quantities discussed in the paper are), the previous equation can
be rewritten as
\be
\hat {\cal O} = {\cal O}\{\peffn,\ptwo(0)\}
\ee
in such a way that the average over $\tau$ is now included in an effective
initial distribution: 
\be
\peffn = \sum_{\tau=0}^{\infty} \Pdep(\tau) \pone(\tau)~.
\label{peff}
\ee

We can make more explicit the physical content of the above reasoning,
that is based on the linearity with respect to the initial distribution of
the two particles.
Atom (2) is deposited with probability $\Pdep(\tau)$ 
a time $\tau$ after atom (1), which means that atom (2)
has the probability $\Pdep(\tau)$  
to find atom (1) distributed according to $\pone(\tau)$:
on average -- and it's now that linearity comes into play -- atom (2)
finds atom (1) with the effective distribution given in Eq.~(\ref{peff}).

In this way we have reduced the problem of evaluating $\hat {\cal O}$
to the evaluation of ${\cal O}$ for two particles deposited
simultaneously. Thus we can ignore the stochasticity of the deposition process
and assume that atoms (1) and (2) land {\it at the same time},
but the actual initial distribution for atom (1) (the uniform
distribution) is replaced by $\peffn$.

The next task is then the determination of $\peffn$.
The function $p_n(\tau)$ (discussed in App.~\ref{ss_dyn})
is the distribution of the first adatom at time $\tau$, i. e., the
solution of the diffusion equation for a single particle with initial
condition
\be
p_n(\tau=0) = \pu_n = {1 \over L^d} ~.
\ee
The sum of $p_n(\tau)$ over all times $\tau$
is the solution of the stationary diffusion equation~(\ref{eq_dif})
(of its discretized version, actually, see App.~\ref{App_rho}), 
whose normalized form will be indicated with $\ps_n$
({\footnotesize S} standing for stationary). 
It has in general
a parabolic form, and in particular in $d=1$, $\ps_n =
[\les L + (L+1) n - n^2]/ [\les L^2 + \hbox{${1\over 6}$} L(L+1)(L+2)]$.

In Appendix \ref{App_eff} it is shown that in all dimensions we can write
\be
\peffn = {\tres \over \tdep + \tres} \ps_n ~.
\label{hatp2}
\ee

The physical content of Eq.~(\ref{hatp2}) is readily understood.
For infinite barriers (regime {\em iii})
$\peffn=\ps_n=1/L^d$: the first particle
cannot escape from the terrace and its distribution is still uniform and
normalized when the second one lands.
For strong but finite barriers (regime {\em ii})
$\peffn =  {\tres \over \tdep} \ps_n = 
{\tres \over \tdep} {1 \over L^d}$: most of the particles
that arrive on the terrace leave it before another particle lands,
but the distribution of the first particle remains practically 
uniform because many attempts are needed to overcome the ES barrier.
In the limit of zero or weak barriers (regime {\em i}) $\peffn =
{\tres \over \tdep} \ps_n$ and $\ps_n$ vanishes on the edges,
reflecting the presence of the absorbing boundaries.

\subsection{The spatial distribution of nucleation events}
\label{spaziotempo}

In the previous subsection we have explained how to transform
the original problem into the new problem of two atoms deposited
at the same time, with  normalized distributions $\ps_n$ (the first)
and $\pu_m$ (the second).
We can now define the probability $R(n,t)$ that
a nucleation event occurs on site $n$ at time $t$
and introduce the following quantities:
\bea
P(n) &=& \sum_t R(n,t) ~ , \\
W &=& \sum_n P(n) ~ . 
\eea

$P(n)$ is the spatial distribution of nucleation events and
$W$ is the probability that two atoms, both on the terrace at time zero,
meet before leaving the terrace. 
It is useful to consider
the {\it normalized} spatial distribution $\Pn(n) = P(n)/W$ as well.

\subsection{The nucleation rate}
\label{ss_omega}

The nucleation rate $\omega$ is defined as the number of
nucleation events per unit time on the {\it whole} terrace of size $L$,
irrespectively of the spatial location of the meeting point.
This quantity is of great importance because it is related
to the probability of second layer nucleation. In a classical
experiment~\cite{class_exp}
a fraction of a monolayer is deposited on the substrate and the
size of islands is made as uniform as possible through an annealing procedure.
Starting from this template a second dose of
atoms is deposited and nucleation on top of existing islands is monitored.
$\omega(L)$ enters in the interpretation of this experiment because
the probability ${\cal P}(t)$ that a nucleation event has occurred on a terrace
by time $t$ is ${\cal P}(t) = 1-\exp\{-\int_0^t d\tau\omega[L(\tau)]\}$.
The rate
$\omega(L)$ is defined and evaluated for a constant terrace size $L$:
in the experiment instead, $L$ grows in time and the time dependence
of $L$ is `system-dependent'. 
Hence,
the growth law $L(\tau)$ of the terrace size must be supplied
beyond $\omega(L)$,
and it depends on the specific morphology of the
surface and the experimental setup.
In other words, the nucleation rate -- on the one hand --
has a very general and basic meaning, but -- on the other hand --
it can hardly be measured directly.
This means that, despite our results for $\omega(L)$ are exact,
the evaluation of $L(t)$ introduces some approximations
in the interpretation of experimental results, whose accuracy depends
on the detail of the system considered.
In addition, some secondary effects, as steering and nonuniform barriers,
may further complicate the problem.

We now connect $\omega$ to $P(n)$ and $W$. In Sec.~\ref{time_scales}
we explained why only processes involving two adatoms are relevant for
studying irreversible nucleation, because $\ttr\ll\tdep$.
If we define the nucleation probability per atom, $\hat\pnuc$, we can
write the nucleation rate as the number of atoms landing on the terrace
per unit time ($FL^d=1/\tdep$) times the nucleation probability per atom:
\be
\omega=FL^d\cdot \hat\pnuc ~~.
\ee

The quantity $\hat\pnuc$ is the probability that a deposited particle 
nucleates a dimer before getting off the terrace and it can be written
as
\be
\hat\pnuc = \sum_{\tau=0}^\infty \Pdep(\tau)
\pnuc \{\pone(\tau),\ptwo(0)\} ~ ,
\label{pnuc}
\ee
where the dependence on the initial distributions of atoms (1) and (2) 
has been made explicit. We stress that the dependence on the initial
distributions occurs {\em via the full diffusion process}. For example,
for independent particles in one dimension, the explicit form of $\pnuc$ is
\be
\pnuc \{\pone(\tau),\ptwo(0) \} = \sum_{m=1}^L \sum_{t=0}^\infty
p_{m,m}(t)\{\pone(\tau),\ptwo(0) \}
\label{pnucexpl}
\ee
where $p_{m,m}(t)$ is the solution of the diffusion equation
{\em in two dimensions} with initial condition given by the product
$p^{(1)}(\tau) p^{(2)}(0)$ (see Ref.~\cite{Politi01b} for more details).

Because of the linearity of $\pnuc$, we have
\be
\hat\pnuc = \pnuc\left\{\peffn,\ptwo(0)\right\} ~.
\label{pnuc2}
\ee

The nucleation probability per atom can be thought as the probability
that atom (1) is still on the terrace when atom (2) is deposited,
times the probability they meet before getting off the terrace.
This is exactly what emerges from Eq.~(\ref{pnuc2}) once
expression~(\ref{hatp2}) for the effective distribution $\peffn$
is inserted:
\be
\hat\pnuc = {\tres \over \tdep + \tres}\cdot \pnuc\{\ps_n,\ptwo(0)\} ~ . 
\ee

The normalization factor of $\peffn$ is the probability that
atom (1) is still on the terrace when the next one shows up;
for infinite ES barriers ($\tres\gg\tdep$) such a probability is 
trivially 1, while for weak and strong barriers
($\tres\ll\tdep$) it is $\tres/\tdep$.
The remaining quantity on the right hand side ($\pnuc\{\ps_n,\ptwo(0)\}$)
is the probability that two atoms, {\it both on the terrace at time zero}
[$\ps_n$ and $\ptwo(0)$ are normalized], meet before descending.
Therefore it coincides with $W$ and we finally obtain
\be
\omega = F L^d {\tres \over \tdep + \tres} W~.
\label{omegagen}
\ee

\subsection{Noninteracting particles}
\label{ss_non}

We are considering a system such that once adatoms come together
an immobile dimer is formed irreversibly:
adatoms stop diffusing and the dimer does not dissociate.
It turns out to be of great help to consider also an artificial model,
with adatoms treated as independently diffusing particles: even if they
meet on the same lattice site they go on diffusing and therefore they
can cross each other several times before leaving the terrace.
We consider all these meetings as `fictitious nucleations',
and define also for noninteracting particles the quantities
mentioned above: the nucleation rate $\oni$, the spatial
distribution $\Pni (n)$ and the total number $\Wni$ of nucleation events,
the subscript {\footnotesize NI} standing for ``Non Interacting".

\section{Equivalence of Mean-Field Theory and noninteracting 
particles model}
\label{mft}

\subsection{The nucleation rate}
\label{nuc_rate}

We have introduced the nucleation rate in Sec.~\ref{ss_omega} and
obtained Eq.~(\ref{omegagen}).
$W$ is the nucleation probability between two atoms that
are both on the terrace at time zero. For the model of noninteracting adatoms
$W$ should be replaced by $\Wni$, the average number of meetings
between the two independent particles. Of course $\Wni$ can be larger than 1.

The simplest and less interesting case is the regime {\em iii} of infinite 
barriers. In such a case $W$ is trivially 1 and $\omega= F L^d = 1/\tdep$,
i. e. any particle deposited on a terrace does form a dimer.
In a sense, this limit is unphysical for mean-field theory because
$\bar\rho$ and $\omf=2dDL^d\bar\rho^2$ diverge when $\les\to\infty$.

In the other two regimes (weak and strong barriers), Eq.~(\ref{omegagen})
becomes
\be
\omega = F L^d \, {\tres \over \tdep} \, W = F L^{2d} \bar\rho W ~,
\label{oex}
\ee
where we have used the relations $\tdep=(FL^d)^{-1}$ and 
$F \tres = \bar\rho$.
We can repeat the same procedure for noninteracting particles and obtain
\be
\oni = F L^d \, {\tres \over \tdep} \, \Wni = F L^{2d} \bar\rho \Wni ~.
\label{oniex}
\ee

It is possible to relate $W$ and $\Wni$ to single-particle
quantities, $\Ndis$ and $\Nall$ (see Sec.~\ref{time_scales}).
They are the number of distinct ($\Ndis$) and all ($\Nall$) sites
visited by a single walker diffusing on the terrace~\cite{nota_UU}.
Let us assume one of the two adatoms fixed on site $s$.
$w(s)$ is the probability that the diffusing adatom visits
site $s$ before getting off the terrace.
$W$ is then the average value of $w(s)$, $W \simeq \sum_s w(s)/L^d$.
The quantity $\sum_s w(s)$ is nothing but the total number of
distinct sites $\Ndis$ visited by the diffusing adatom, so that
$W\simeq\Ndis/L^d$.
The same argument for noninteracting particles gives 
$\Wni\simeq\Nall/L^d$.

\begin{figure}
\centerline{\psfig{figure=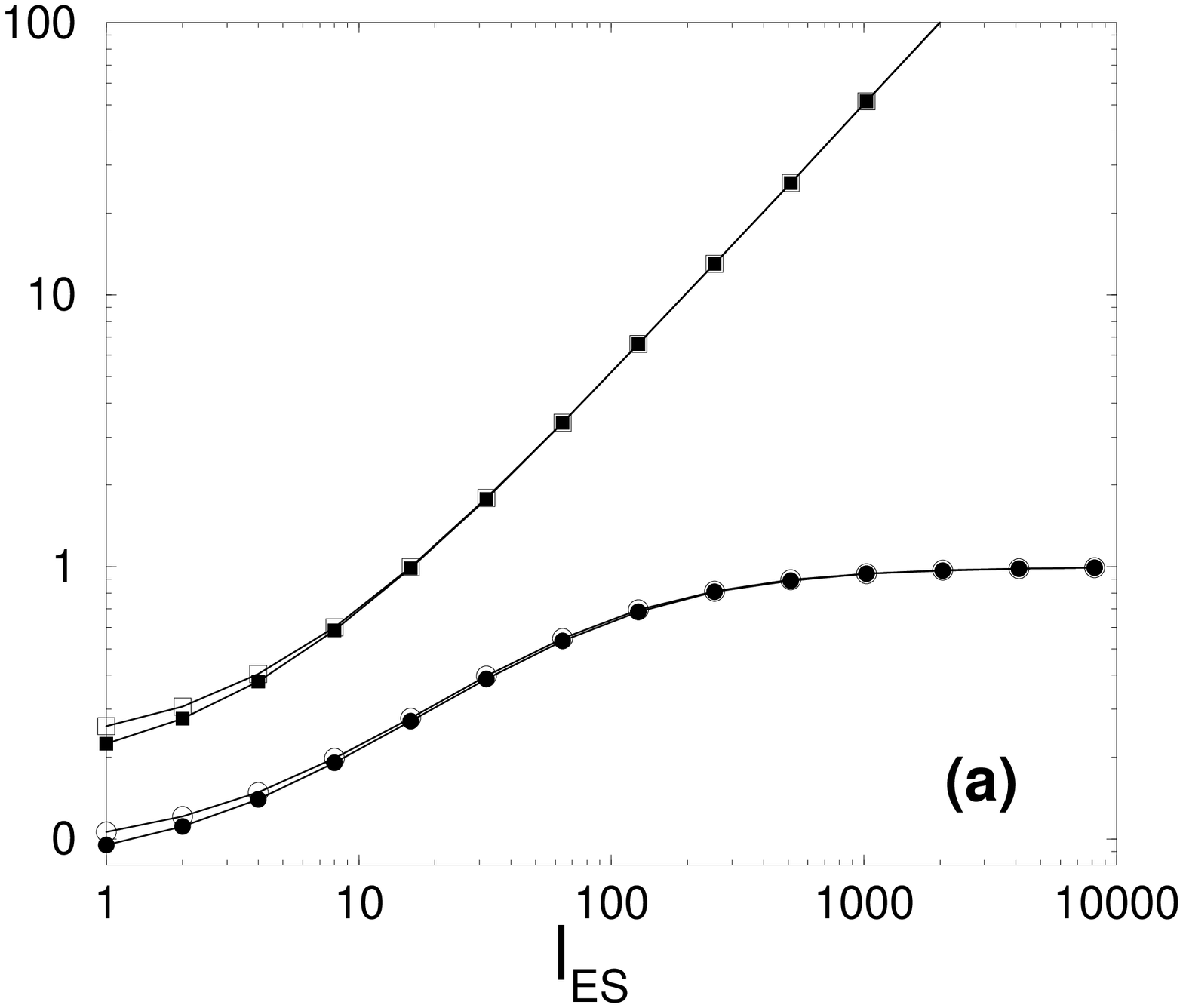,width=8cm,angle=-90}
\psfig{figure=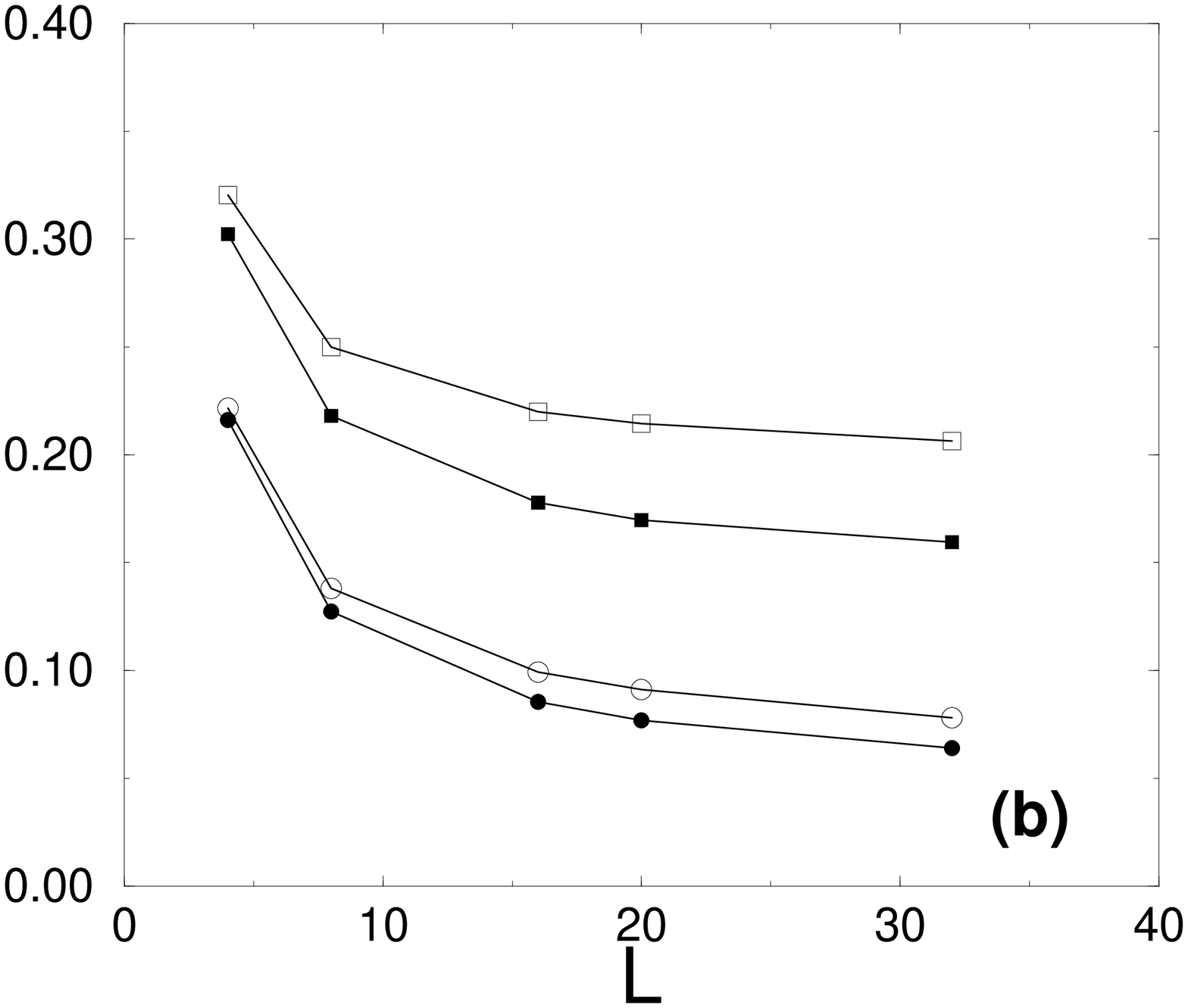,width=8cm,angle=-90}}
\caption{(a) Log-log plot of $W$ (empty circles), $\Ndis/L^2$
(full circles), $\Wni$ (empty squares), $\Nall/L^2$ (full squares)
versus $\les$ for $d=2$ and $L=20$.
(b) The same quantities plotted versus $L$ for $d=2$ and $\les=0$.}
\label{fig:WandN}
\end{figure}

The relations:
\be
W\simeq {\Ndis\over L^d}  \hbox{~~~~and~~~~} \Wni\simeq {\Nall\over L^d}
\label{eq:WandN}
\ee
have been derived under the assumption that one atom is immobile.
In Fig.~\ref{fig:WandN} we compare numerically the values
of single-particles quantities ($\fra{\Ndis}{L^d},\fra{\Nall}{L^d}$)
with two-particles quantities ($W,\Wni$).
The former have been calculated via Monte~Carlo simulations and the latter
through the numerical solution of the discrete diffusion equation for
two atoms on a terrace (discussed in detail in the next article):
it comes out that relations (\ref{eq:WandN}) are
well satisfied, so that assuming one atom as immobile is perfectly
reasonable for the evaluation of $W$ and $\Wni$.

If we insert the relations (\ref{eq:WandN}) into
Eqs.~(\ref{oex},\ref{oniex}) we obtain:
\bea
\omega &\simeq& FL^d\bar\rho\Ndis \\
\oni &\simeq& FL^d\bar\rho\Nall ~ .
\eea

Since $\Nall$ is related to the residence time by $\tau_{res} = 
\Nall/(2dD)$, we can write 
\be
\oni \simeq 2dF L^d \bar\rho D \tres = 2dD L^d {\bar \rho}^2 = \omf ~ .
\ee

In this way we have shown that for the nucleation rate {\it the mean-field
treatment is equivalent to considering particles as noninteracting}, i. e.
counting also meeting events following the first one.
For this reason the mean-field value is an
overestimate of the correct nucleation rate.
Furthermore we have proven that
\be
{\omf \over \omega} \simeq
{\Wni \over W} \simeq
{\Nall \over \Ndis} \equiv {\cal N} ~ .
\label{corrfact}
\ee

\begin{figure}
\centerline{\psfig{figure=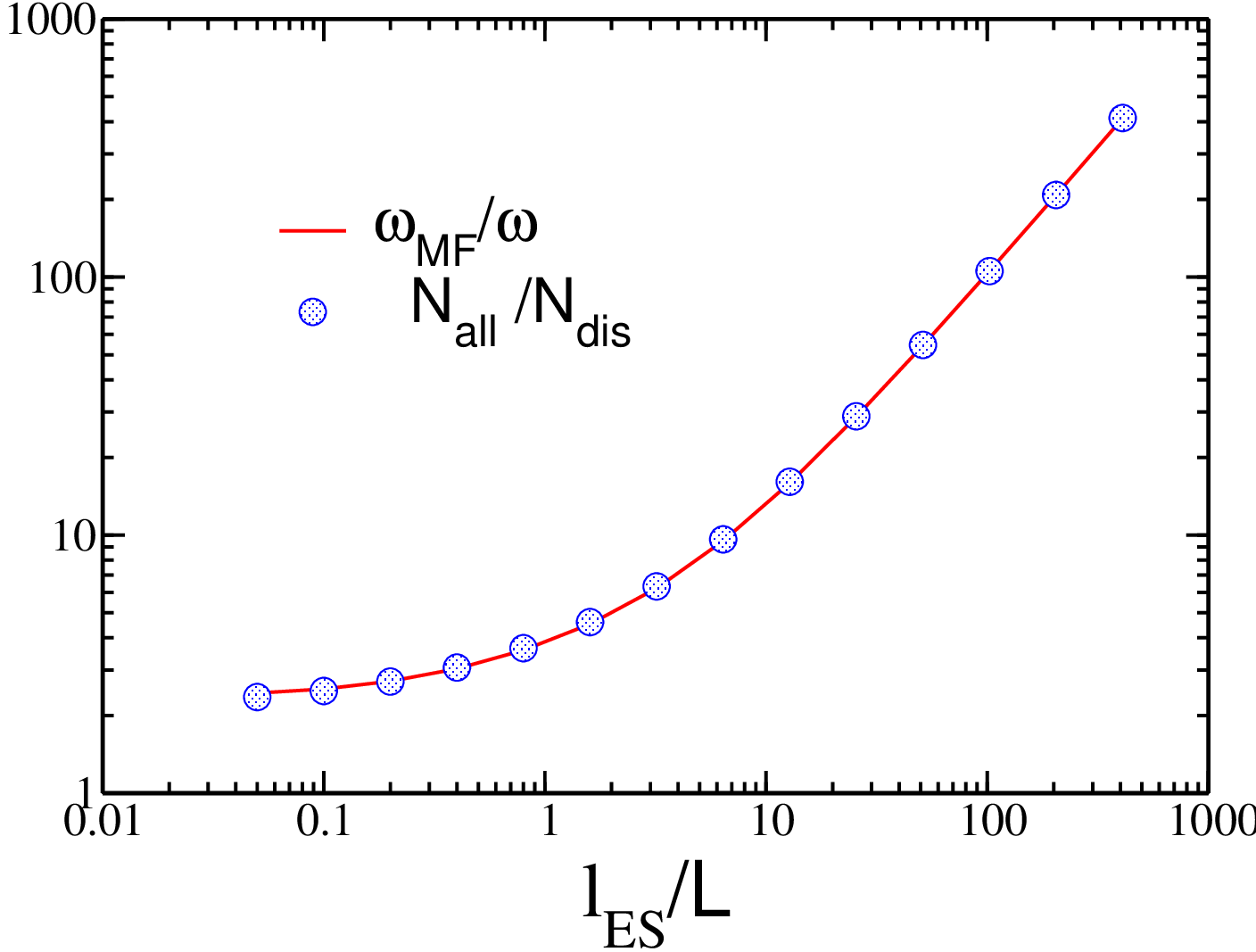,width=8cm,angle=-90}}
\caption{Plot of the correction factor $\omf / \omega$ 
and of $\cal N \equiv \Nall/ \Ndis$ versus $\les/L$,
for a square terrace of size $L=20$.}
\label{fighighlights}
\end{figure}

In Fig.~\ref{fighighlights}, the comparison of $\omf / \omega$,
computed exactly in the companion paper~\cite{Politi01b} with
$\Nall /\Ndis$, evaluated numerically, shows clearly that
Eq.~(\ref{corrfact}) is valid with great accuracy.

The correction factor $\cal N$ depends on well known
properties of single particles performing a random walk.
The numerator $\Nall$ is just (see Sec.~\ref{time_scales}): $\Nall =
2dL(\beta L+\alpha\ell_{\hbox{\tiny ES}})$. 
The value of the denominator $\Ndis$ is well known~\cite{Hughes} in the
absence of step-edge barriers, being of order $L$ in $d=1$ and of order
$L^2/\ln L$ in $d=2$, and it is trivial in the limit of infinite barriers,
being exactly equal to $L^d$. 
Hence in $d=1$ we obtain ${\cal N}\sim (L+\alpha\ell_{\hbox{\tiny ES}})$,
for all $\les$.

\begin{figure}
\centerline{\psfig{figure=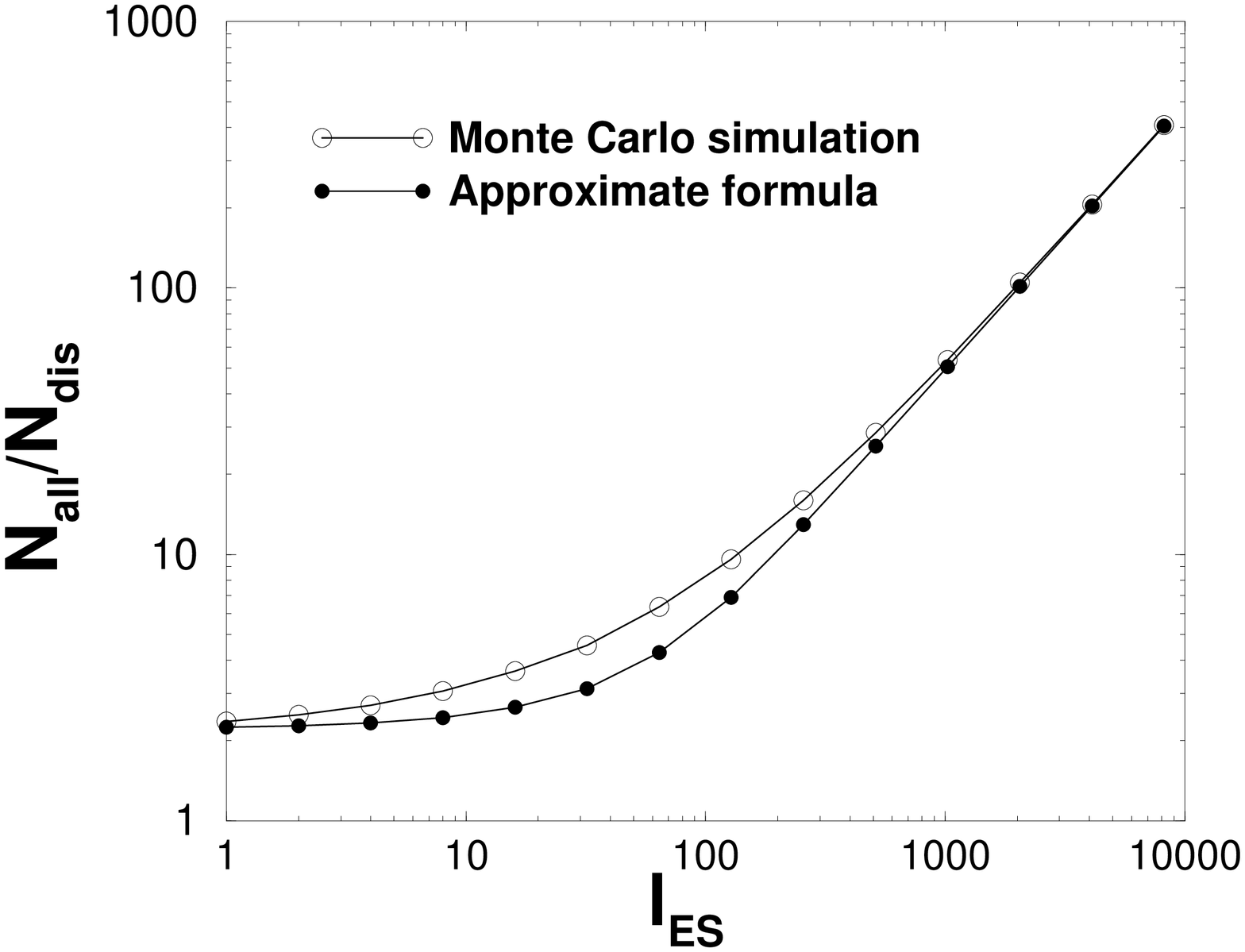,width=8cm,angle=-90}}
\caption{Log-log plot of $\Nall/\Ndis$ versus $\les$ for $d=2$ and $L=20$.}
\label{fig:N}
\end{figure}

In $d=2$ we have the limiting expressions ${\cal N}\sim\ln L$ for weak
barriers and ${\cal N}\sim \les/L$ for strong ones.
For intermediate barriers it is possible to find a simple approximate
expression for $\Ndis$ and therefore
an interpolation between the two limits.
The atom performs on average a number $N_{tr}=\tres/\ttr$
of traversals of the island. During a single traversal each site
has a probability $p_1\sim (1/\ln L)$ to be visited. After all 
$N_{tr}$ traversals the probability $p_s$ that a generic site has been
visited at least once is given by $1-p_s = (1-p_1)^{N_{tr}}$. Hence
we can estimate the number of distinct visited sites as
\be
\Ndis = L^2 p_s = L^2 \left[1-(1-p_1)^{\tres/\ttr}\right] ~ .
\ee
This expression assumes all traversals to be independent,
which is clearly not strictly correct. However it gives the right values
in the limits $\les=0$ and $\les=\infty$ and for intermediate
barriers its accuracy can be tested numerically.
In Fig.~\ref{fig:N} we have plotted the ratio $\Nall/\Ndis$ as
a function of $\les$, for $L=20$.
The picture shows a reasonable agreement between the analytical
estimate and the numerical simulation.

\subsection{The spatial distribution}
\label{spat_dist}

In the previous subsection we have shown that MF theory
overestimates the nucleation rate by the quantity
${\cal N}$ because it counts all meetings between
two noninteracting adatoms.
We are now going to prove that the identification of the mean-field
approach as a theory of noninteracting particles is valid for the 
spatial distribution of nucleation events as well.
We adopt a continuum notation so that a single proof is sufficient to
demonstrate that $\Pni(x)$ and $\rho^2(x)$ are proportional in any
dimension, for any value of the ES barrier and for any terrace shape.
In the regimes of strong and infinite ES barrier this result is
trivial because both $\rho^2(x)$ and $\Pni(x)$ are constant.

We face the problem of determining the
quantity $\Pni(x)$
for a pair of adatoms, one with initial distribution
$p^{(2)}(x,0)=\pu(x)$ and the other with the effective distribution
$p^{(1)}(x,0)=\peffx=\sum_t p^{(2)}(x,t)=\rho(x)$.

We can consider the coordinates $(x_1,x_2)$ of the two atoms
($x_{1,2}$ are vectors in a $d$-dimensional space) as defining
the position $\bold{x}=(x_1,x_2)$ of a single particle in a space
of dimensionality $d'=2d$.
This particle moves according to the diffusion equation
$\partial_t p=\fra{D}{2}\nabla^2 p$.
The factor $\fra{1}{2}$ appears because of the different time step
$\Delta t$ employed in describing a single walker 
($\Delta t=1/2dD$) or two walkers ($\Delta t=1/2d'D$)  on a terrace.

Integrating in time and defining $P(\bold{x})=\int_0^\infty dt p(\bold{x},t)$
one finds
\be
\fra{D}{2} \nabla^2 P(\bold{r}) = -p(\bold{x},0) ~ .
\ee

An interchange of the two particles 
[$p^{(1)}(x,0)=\pu(x)$ and $p^{(2)}(x,0)=\rho(x)$] is equally legitimate and
it is useful to use a simmetrized form for $p(\bold{x},0)$:
\be
p(\bold{x},0)
= \fra{1}{2} \left[\rho(x_1)\pu(x_2) + \pu(x_1)\rho(x_2) \right] ~ .
\ee
Notice that $\rho(x)$ is also the solution of the equation
$D\nabla^2\rho(x)= - \pu(x)$. Therefore
\be
\fra{D}{2} \nabla^2 P(\bold{r}) = \fra{D}{2}
[\rho(x_1)\nabla^2_2\rho(x_2) + \rho(x_2)\nabla^2_1\rho(x_1) ] ~ ,
\ee
where $\nabla^2_i$ acts on $x_i$ only and $\nabla^2\equiv
\nabla^2_1 + \nabla^2_2$. Hence
\be
\nabla^2 P(\bold{r}) = \nabla^2 [\rho(x_1)\rho(x_2)] ~ ,
\ee
i. e. the function $\chi(\bold{r}) = P(\bold{r}) 
- \rho(x_1)\rho(x_2)$
is harmonic. It is easy to
show~\cite{nota:arm} that $\chi(\bold{r})$ must be identically zero.
Hence:
\be
P(\bold{r}) = \rho(x_1)\rho(x_2) ~ .
\ee 

If we set $x_1=x_2=x$, the left hand side is just the nucleation
probability $\Pni(x)$ at point $x$ between two noninteracting adatoms,
and the right hand side is the mean-field prediction.
Notice that we have not used the explicit form of $\pu(x)$.
Hence, the proof holds for {\it any}
initial spatial distribution $\pu(x)$, so that the equivalence
between MF theory and the noninteracting particles model is true even if
atoms are not deposited uniformly. 

So far we have rigorously shown that the MF result for $P(n)$
is not exact.
However, one may wonder whether the error introduced by taking into
account all meeting events following the first one is 
expected to be large or negligible.
We address this issue by evaluating the relative
weight of successive encounters for noninteracting particles.

Let us consider noninteracting particles and
define $\mu_j$ as the fraction of times the $j$-th meeting
event actually occurs. Clearly $\mu_0=1$ and $\mu_j \geq \mu_{j+1}$.
Let us also define the {\em normalized} distribution for the $j$-th
nucleation event $\PNI_j(n)$.
Notice that $\PNI_j(n)=P^{(N)}(n)$, the distribution for interacting
particles.

The total distribution of nucleation sites is simply
\be
\Pni(n) = \sum_{j=1}^\infty \mu_j \PNI_j(n)
\label{rho}
\ee
and the quantities $W$ and $\Wni$ are given by:
\bea
W &=& \mu_1 ~ , \\
\Wni &=& \sum_n \Pni(n) = \sum_{j=1}^\infty \mu_j ~ .
\eea

If we now introduce the normalized distribution
of all fictitious nucleations following the first one:
\be
\hat\PNI(n) = {\sum_{j=2}^\infty \mu_j \PNI_j(n)
\over \sum_{j=2}^\infty \mu_j } ~ ,
\ee
we can write
\be
\Pni(n) = \mu_1 \PNI_1(n) + \sum_{j=2}^\infty \mu_j \PNI_j(n)=
W P^{(N)}(n) + \hat\PNI(n) (\Wni-W) ~ .
\ee
For weak barriers, in $d=1$, the weight $W$ of the first term is
constant, while the second one diverges as $L$. As a consequence,
for large $L$ the distribution $\Pni(n)$ is dominated by the contribution of 
the fictitious successive nucleations.
In $d=2$ the first term goes as $1/\ln L$ while the second is constant.
Again, for large $L$, the contribution of first nucleation events
becomes negligible.
For strong and infinite barriers, $W=1$ while $\Wni$ is infinite, so
$\Pni(n)$ coincides with $\PNI_\infty(n)$.

In all cases the MF expression for the spatial
distribution of nucleation sites [$\Pni(n)$] is dominated for large $L$
by the contribution of the fictitious nucleations following the first one. 
A priori there is no reason for supposing that
the distribution of the $j$-th nucleation event is
equal to the distribution of the first one, so we expect that
the difference between the MF spatial distribution and the exact result
persists for all values of $L$.
This will be checked and confirmed in the following paper~\cite{Politi01b}.

\section{Conclusions}
\label{conc}

This paper has been devoted to an accurate investigation of the mean-field
approach to the problem of irreversible nucleation.
The main outcome is the proof that MFT
is equivalent to a model where particles do not interact and all
their meetings are counted as fictitious nucleations.

In the regime of infinite ES barriers, MFT simply breaks down because
it predicts a diverging nucleation rate, in contrast to the correct
value $\omega=FL^d$. In the other, physically more interesting,
regimes the equivalence of MFT with the model of noninteracting particles
implies that $\omf$ overestimates the correct nucleation
rate by the factor ${\cal N}=\Nall /\Ndis$.
This ratio has a clear meaning: a diffusing adatom passes ${\cal N}$
times on a visited site. 
It depends on single-particle quantities ($\Nall,\Ndis$) whose
expressions are well known from the theory of random walks.

In Table~\ref{tabella} we summarize the value of the correction factor
${\cal N}$ in regimes {\it i} and {\it ii} and we report the 
approximate expressions for the nucleation rate $\omega$.
They are approximate in the sense that numerical prefactors are
neglected, but they scale correctly with $L,D,D',a_0$.
The lattice constant $a_0$ has been reintroduced in order to
give dimensionally correct espressions.
Also, we have made explicit the dependence of $\les$ on $D$
and $D'$, so that only basic quantities appear.

\begin{table}
\begin{tabular}{c|c|c|c|c}
 & \multicolumn{2}{c|}{({\it i})} & \multicolumn{2}{c}{({\it ii})}  \\
 & \multicolumn{2}{c|}{${D-D'\over D'}\ll {L\over a_0}$}
 & \multicolumn{2}{c}{${D\over D'}\gg {L\over a_0}$ ~~~ 
$D'\gg FL^{d+1}a_0$} \\ \hline
 & $\omega$ & ${\cal N}$
 & $\omega$ & ${\cal N}$  \\ \hline
$~~d=1~~$ & ${F^2L^4\over D}$ & ${L\over a_0}$ &
        ${F^2L^3a_0\over D'}$ & ${D\over D'}$  \\
$~~d=2~~$ & ${F^2L^6\over D\ln(L/a_0)}$ & $\ln(L/a_0)$ &
        ${F^2L^5a_0\over D'}$ & ${Da_0\over D'L}$  \\
\end{tabular}
\caption{We report the nucleation rate $\omega$ and the
correction factor ${\cal N}$ in the two
regimes of weak ({\it i}) and strong ({\it ii}) step-edge barriers
and in one and two dimensions. The conditions defining regimes
{\it i} ($\tres\simeq\ttr$) and {\it ii} ($\ttr\ll\tres\ll\tdep$) 
are written in terms of the basic quantities $L,D,D',a_0$.}
\label{tabella}
\end{table}

The expression $\omega\sim F^2L^5a_0/D'$, valid in two dimensions
for strong step-edge barriers has already been given in
Ref.~\cite{KPM}. It is worth repeating that
the nucleation rate in this limit does not depend on the diffusion
constant $D$ so that the nucleation rate can not be promoted 
by using surfactants.

Application of MFT is acceptable only in the regime of vanishing barriers 
in two dimensions, because in this case the correction factor 
[${\cal N}=\ln(L/a_0)$] is a small number, for realistic terrace sizes.

In order to obtain exact expressions for $\omega$ it is necessary to
have an accurate estimate of $W$, or equivalently of $\Ndis$.
$W$ is a function of the terrace size $L$ and of the ES length $\les$: 
for strong barriers $W=1$, while
for weak barriers $W\sim 1$ in $d=1$ and $W\sim 1/\ln L$ in $d=2$.
So, for realistic values of $L$, $W$ depends on $L$ and $\les$
much more weakly than the other quantities appearing in
$\omega=FL^{2d}\bar\rho W$.
However, its dependence is not fully
negligible: Figs.~\ref{fig:WandN}a and~b show that (in $d=2$) for
$L=20$, $W$ varies by a factor~10 by changing $\les$ from zero
to infinity and for $\les=0$, $W$ varies by a factor~3 
by changing the terrace size from $L=4$ to $L=32$
($\ln 32/\ln 4=2.5$).
For comparison, the quantity $\bar\rho$, which appears along with $W$
in the expression for $\omega$, varies by a factor 50 by
changing $\les/L$ from 0 to 6 and by a factor 64 by changing $L$
from 4 to 32, for $\les=0$.
The problem of the exact determination of $W$ will be tackled
again in the following paper.

A last comment on rate equations and mean-field approximation is in
order here. According to the former, the nucleation rate is written
$\ore=\sigma_1 D\rho^2$ and the latter corresponds to taking $\sigma_1$
as a constant. In general, $\sigma_1$ is defined through the
relation $\Phi_1=D\sigma_1\rho$, where $\Phi_1$ is the flux of
atoms attaching to an adatom.
The resulting relation $\ore=\Phi_1\rho$ is {\it exact}, if
$\Phi_1$ is evaluated correctly; for example, we can solve the diffusion
equation for a single walker on the terrace, where a second walker is taken 
as an absorbing sink. Since we have shown that the nucleation rate can be
evaluated assuming an atom as immobile, such treatment is
essentially correct.
In other words, if the capture number is not taken as a constant, the
expression $\ore=\sigma_1D\rho^2$ may give correct results,
but this method has nothing to do with the usual mean-field approach.

Finally, with regard to the spatial distribution, we have provided
a very general demonstration of the equivalence between mean-field 
and noninteracting particles.
We have also shown that the difference between
$P(n)$ and $\Pni(n)$ is not an effect of the finite size of the
terrace and it remains true for large $L$.
The full computation of the spatial distribution of nucleation events
requires the solution of the dynamical problem of two interacting
atoms diffusing on a terrace. This problem will be solved
analytically and/or numerically in the following paper.

\acknowledgements
Authors gratefully acknowledge useful discussions and correspondance
with Jim Evans, Thomas Michely and Filippo Colomo.
                 
\appendix

\section{Single particle on a terrace}
\label{one_par}
In this Appendix we summarize the behavior of a single particle on a
terrace for all values of the Ehrlich-Schwoebel length $\les$.

\subsection{The stationary adatom density}
\label{App_rho}

The discrete evolution equation for a particle in a cubic $d$-dimensional
space is
\be
p_n(t+1) = {1\over 2d} \sum_\delta p_{n+\delta}(t) ~ ,
\label{evo_dis}
\ee  
where $n+\delta$ indicates a neighbour of the site $n$.
If we sum over $t$ and define the quantity 
$\rho_n = \sum_{t=0}^\infty p_n(t)$ we obtain
\be
\left[ \sum_\delta \rho_{n+\delta} - 2d \rho_n \right] + 2d p_n(0) = 0 ~ .
\label{eq:stat}
\ee

The terms in square brackets give the discrete laplacian of $\rho_n$;
therefore the sum $\rho_n=\sum_t p_n(t)$ is simply
the solution of the stationary diffusion equation in the presence of
the flux $2d p_n(0)$.

In $d=1$, for constant $p_n(0)$,
it is possible to find the exact discrete solution for any value
of $\les$, once we remark that the general solution of the 
homogeneous equation is $\rho_n=c_0+c_1n$ and a particular solution
of the nonhomogeneous equation is $\rho_n=-c_2n^2$ [the factor $c_2$
depending on the constant term in Eq.~(\ref{eq:stat})].

Boundary conditions are $\rho_0=a\rho_1$ and $\rho_{L+1}=a\rho_L$, where
$a=\les/(1+\les)$ goes from 0 to 1 as 
the Ehrlich-Schwoebel length $\les$ varies from 0 to $\infty$.

The explicit expression of $\rho_n$ is
\be
\rho_n = {1 \over L} [\les L +(L+1)n - n^2]~.
\ee
Its normalized version is
\be
\ps_n \equiv {\rho_n \over \sum_n \rho_n}
= { 1\over \les L^2 + \fra{L(L+1)(L+2)}{6} } [ \les L +(L+1)n - n^2 ] ~ .
\label{disc_1d}
\ee
In a continuum formalism, the equation is
$D\partial_x^2 \rho +F =0$ and the solution in $d=1$ is
\be
\rho(x) = \fra{F}{2D} [\les L + Lx -x^2 ] ~ .
\label{cont_1d}
\ee

In $d=2$ the solution of the continuum equation is as easy as in $d=1$
if we specialize to a circular terrace. If $L$ now denotes the radius,
the solution for generic $\les$ is:
\be
\rho(r) = \fra{F}{4D} [ L^2 + 2L\les - r^2 ] ~ .
\label{cont_2d}
\ee
We finally evaluate the average density $\bar\rho$ on the terrace:
\bea
\bar\rho &=& \fra{F}{12D} L(L+6\les) ~~~d=1 \label{rho_1d}\\
\bar\rho &=& \fra{F}{8D} L(L+4\les)  ~~~d=2 ~~~
\hbox{[circular terrace]}
\label{rho_2d}
\eea

\subsection{The dynamical problem in one dimension}
\label{ss_dyn}

We now summarize the dynamical behavior of a single particle
on a one-dimensional terrace. The two-dimensional case is treated 
in the next Subsection.

The discrete evolution equation for the particle is
\be
p_n(t+1) = {1\over 2} [p_{n+1}(t) + p_{n-1}(t)] ~ ,
\label{dif1d}
\ee
with the usual boundary conditions,
$p_0(t)= a p_1(t)$ and $p_{L+1}(t)= a p_L(t)$.
The solution is found by separating the space and time variables, 
$p_n(t)=X(n) F(t)$:
\be
{F(t+1)\over F(t)} = \lambda = {X(n+1)+X(n-1)\over 2
X(n)} ~~~~0<\lambda <1 ~ .
\ee
The temporal part is $F(t)= \lambda^t F(0)$.
The spatial part has the general form
\be
X(n)=A \sin(n\phi)+B \cos(n\phi) ~ ,
\ee
which gives $\lambda=\cos\phi$. The boundary conditions determine
the values of $A$, $B$ and $\phi$.

In particular, by imposing the boundary condition in $n=0$ one obtains
$B=bA$ with $b=a \sin\phi/(1-a\cos\phi)$.
Using this relation and imposing the other boundary condition
in $n=L+1$ one obtains
\be
\tan(L\phi)={(a^2-1) \sin\phi \over (1+a^2) \cos\phi-2 a} ~ .
\label{genericaphi}
\ee

This equation has $L$ solutions which we label as $\phi_k$ with
$k=1,\dots,L$. Then the general solution is
\be
p_n(t)=\sum_{k=1}^{L} 
\cos^t \left(\phi_k \right) X_k(n) ~ ,
\label{generica}
\ee
with
\be
X_k(n) = A_k \sin(\phi_k n) + B_k \cos(\phi_k n) ~ .
\ee

Given $p_n(t)$ one can compute $S(t)$,
the probability that an adatom is still on the terrace at time
$t$ after deposition (survival probability):
\be
S(t) \equiv \sum_{n=1}^L p_n(t) ~ .
\label{Sdit}
\ee
Another important quantity is the residence time, defined as
\be
\tres \equiv \sum_{t=1}^\infty t[S(t-1)-S(t)]
\ee
because $S(t-1)-S(t)$ is the probability that the
particle stays on the terrace exactly a time $t$.
It is easy to check that
\be
\tres = \sum_{t=0}^\infty S(t) = 
        \sum_{n=1}^L \left[ \sum_{t=0}^\infty p_n(t) \right] =
        \sum_{n=1}^L \rho_n ~ .
\label{tres}
\ee
Recalling App.~\ref{App_rho}, for the initial distribution
$p_n(0)= \pu_n=1/L$ we have, for any value of the ES barrier,
\be
\tres = {(L+1)(L+2)\over 6} + \les L ~ .
\label{tresgeneric}
\ee

In order to pass to a continuous time we have to multiply it by
$\Delta t=1/2D$. For large $L$, $\tres=\fra{L}{D}
(\fra{L}{12}+\fra{\les}{2})$. This result agrees with the
relation $\bar\rho=F\tres$ [see Eq.~(\ref{rho_1d})].

Unfortunately it is not possible to solve explicitly Eq.~(\ref{genericaphi})
for generic values of $a$:
we now discuss the two limit cases where an explicit
solution is possible.

\subsubsection{Zero barriers}
For $\les=0$ ($a=0$), the allowed values of $\phi_k$ are
\be
\phi_k={ k \pi \over L+1} ~, ~~~~~~~~ (k=1,\dots,L)
\ee
and the general solution is
\be
p_n(t)=\sum_{k=1}^L A_k \cos^t \left({k\pi\over L+1}\right)
\sin\left({n k\pi\over L+1}\right) 
\label{sol1d}
\ee
with
\be
A_k = {2\over L+1} \sum_{n=1}^L p_n(0) \sin\left({n k\pi\over L+1}\right) ~ .
\ee

In particular, two forms of $p_n(0)$ are most interesting to us.
For a uniform distribution $p_n(0)=\pu_n=1/L$ the coefficients are
\be
A_k \equiv A_k^U = {2\over L(L+1)} \sin\left({L\over 2}{k\pi\over L+1}\right)
\sin\left({k \pi\over 2}\right)
\csc\left({1\over 2}{k\pi\over L+1}\right) ~ .
\label{AkU}
\ee
For the distribution $p_n(0)=\ps_n = {6 \over L(L+1) (L+2)} n(L+1-n)$
[see Eq.~(\ref{disc_1d})], the explicit solution is:
\be
A_k \equiv A_k^S = {6 \over L (L+1)^2 (L+2)}
{\sin\left({k \pi \over 2}\right) \over 
\sin^3\left[{k \pi \over 2(L+1)}\right]}
\sin\left[{L k \pi \over 2(L+1)}\right] ~ .
\label{AkS}
\ee

As shown in App.~\ref{App_rho}, $\ps_n$ is the
normalized version of $\rho_n=\sum_{\tau=0}^\infty p_n(\tau)$ where
$p_n(\tau)$ is the solution of the diffusion equation with uniform
initial condition $\pu_n$.
Writing explicitly the sum we obtain
\bea
\sum_{\tau=0}^\infty p_n(\tau) & = & \sum_{\tau=0}^\infty 
\sum_{k=1}^L A_k^U \cos^{\tau} \left({k\pi\over L+1}\right)
\sin\left({n k\pi\over L+1}\right) \\
& = & \sum_{k=1}^L {A_k^U \over 1-\cos \left({k\pi\over L+1}\right)}
\sin\left({n k\pi\over L+1}\right) ~ .
\eea
Hence 
\be
A_k^S \propto {A_k^U \over 1-\cos \left({k\pi\over L+1}\right)} =
{A_k^U \over 2 \sin^2 \left[{k\pi\over 2(L+1)}\right]} ~ ,
\ee
as can be easily verified by comparing Eq.~(\ref{AkU}) with~(\ref{AkS}).

If we sum $p_n(t)$ over $n$ [see Eq.~(\ref{sol1d})] we obtain
the survival probability:
\be
S(t) = \sum_{k=1}^L  A_k \cos^t \left({k\pi\over L+1}\right)
\sin\left({L\over 2}{k\pi\over L+1}\right) \sin\left({k \pi\over 2}\right)
\csc\left({1\over 2}{k\pi\over L+1}\right) ~ .
\ee

The distribution $p_n(t)$ is in general a sum of exponential decays
\be
p_n(t) = \sum_{k=1}^L A_k \sin\left({n k\pi\over L+1}\right)
\exp\left[t \ln \cos \left({k\pi\over L+1}\right)\right] ~ .
\ee
It can be considered as a single exponential when the second slowest
decaying exponential is negligible. For $L \gg 1$ this means
\be
\exp\left[-\left({2 \pi \over L} \right)^2 {t \over 2} \right] \ll 1
~~~ \Rightarrow ~~~ t \gg {L^2 \over 2 \pi^2} \simeq \ttr ~ .
\ee
Hence, for $t \gg \ttr$
\be
p_n(t) \simeq A_1 \sin\left({n \pi \over L+1} \right)
\exp \left[- \left({\pi \over L}\right)^2 {t \over 2} \right]
\simeq A_1 \sin \left(n \pi \over L+1 \right)
\exp \left(- {t \over \tres} \right) ~ .
\ee
For the same reason, for $t \gg \ttr$
\be
S(t) \simeq A_1 {2(L+1) \over \pi} \exp\left(- {t \over \tres}\right) ~ .
\ee

\subsubsection{Infinite barriers}

For $\les=\infty$ ($a=1$) the allowed values of $\phi_k$ are
\be
\phi_k={ k \pi \over L} ~~~~~~~~ (k=0,\dots,L-1)
\ee
and $A_k = B_k \tan\left({k \pi \over 2 L}\right)$,
so that the general solution is 
\be
p_n(t)=\sum_{k=0}^{L-1} A_k  \cos^t \left({k\pi\over L}\right) X_k(n) ~ ,
\ee
where
\be
X_k(n) = \left[\tan\left({k \pi \over 2 L}\right)
\sin\left({n k\pi\over L}\right) +\cos\left({n k\pi\over L}\right) \right] ~ .
\ee

The coefficients $A_k$ depend on the initial condition
through the relation:
\be
A_k= {1 \over N_k} \sum_{n=1}^L p_n(0) X_k(n) ~ ,
\ee
where ($\delta_{k0}$ is the Kronecker symbol):
\be
N_k = {L \over 2}\left[1 +\tan^2\left({k \pi \over 2 L}\right)  \right]
(1 + \delta_{k0})~.
\ee

$p_n(t)$ is the sum of a constant (the term for $k=0$)
and exponentially decaying terms ($k>0$).
For $p_n(0)=1/L$, the only nonvanishing coefficient is
$A_0 = 1/L$ and this implies for all times
\be
p_n(t) = {1 \over L} ~ .
\ee

In the general case of nonconstant $p_n(0)$, the exponential decays are 
negligible when $\exp\left[- \left({\pi \over L}\right)^2 {t \over 2}
\right] \ll 1$, that is to say $t \gg {2 \over \pi^2} L^2 \simeq \ttr$.

\subsubsection{Strong barriers}
Let us consider now the case of finite but large $\les$ ($\les \gg L$).
The solution of Eq.~(\ref{genericaphi}), with $a \to 1$ and large but
fixed $L$, yields, for the two smallest $\phi_k$
\bea
\phi_1 &\simeq & \sqrt{{2 (1-a) \over L}} = \sqrt{{2 \over L \les}} ~ , \\
\phi_2 &=& {\pi \over L} + O(1-a) .
\eea

The slowest decays in the general solution are
therefore $\exp(-\phi_1^2 t/2)$ and $\exp(-\phi_2^2 t/2)$.
For finite values of $L$ we can neglect the second exponential for times
such that
\be
\exp\left[t \ln \cos (\phi_2) \right] \ll 1
~~~ \Rightarrow ~~~ t \gg {2 \over \pi^2} L^2 \simeq \ttr ~ .
\ee

Hence for times larger than $\ttr$ one can write
\be
p_n(t) = B_1 \cos\left(n \sqrt{{2 \over L \les}} \right)
\cos^t \left(\sqrt{{2 \over L \les}} \right) \simeq
B_1 \exp \left(-{t \over L\les} \right) ~ ,
\label{eq:pnt}
\ee
where $B_1 \simeq 1/L$ and $\tres=L \les$. This value of $\tres$,
multiplied by $\Delta t= 1/2D$ coincides with its continuum counterpart
$\bar \rho/F = L \les /(2D)$.

\subsection{The dynamical problem in two dimensions}

It is useful to summarize here some results for a single particle
on a two-dimensional terrace. The general solution is

\be
p_{m,n}(t) = \sum_{k,j=1}^L A_{kj}{1\over 2^t}\left[
\cos\left({k\pi\over L+1}\right) + \cos\left({j\pi\over L+1}\right)\right]^t
X_k(m) X_j(n) ~ ,
\label{general}
\ee
where the coefficients $A_{kj}$ are
\be
A_{kj} = {1 \over N_k N_j} \sum_{m,n=1}^L p_{m,n}(0)
X_k(m) X_j(n) ~ .
\ee

For zero barriers $X_k(n) = \sin \left({n k \pi \over L+1}\right)$
and $N_k = {L+1 \over 2}$.
For a uniformly distributed adatom, 
$\pu_{m,n}=1/L^2$ and the coefficients are
\bea
A_{kj}^U = A_k^U A_j^U = && \left[ {2 \over L (L+1)} \right]^2
\sin\left({k \pi \over 2} \right) \sin\left({j \pi \over 2} \right)
\sin\left[{L k \pi \over 2(L+1)} \right] \\
&\times & \sin\left[{L j \pi \over 2(L+1)} \right] 
\csc\left[{k \pi \over 2(L+1)} \right]
\csc\left[{j \pi \over 2(L+1)} \right] ~ .
\eea

We indicate as $\ps_{m,n}$ the normalized solution of the stationary
diffusion equation in the presence of a constant flux.
Differently from what occurs in the one-dimensional case, the explicit
form of $\ps_{m,n}$ is not known exactly for a square terrace. 
However, the expression of its coefficients
$A_{kj}^S$ can be obtained by exploiting the property (see
App.~\ref{App_rho}) that $\ps_{m,n} = N \sum_{\tau=0}^\infty
p_{m,n}(\tau)$ where $p_{m,n}(\tau)$ is the solution of the diffusion
equation with uniform initial condition $\pu_{m,n}=1/L^2$ and
$N = 1/\tres$ is a normalization factor:
\bea \nonumber
\sum_{\tau=0}^\infty p_{m,n}(\tau) & = & \sum_{\tau=0}^\infty
\sum_{k,j=1}^L A_{kj}{1\over 2^\tau}\left[
\cos\left({k\pi\over L+1}\right) +
\cos\left({j\pi\over L+1}\right)\right]^\tau
\sin\left({m k\pi\over L+1}\right)
\sin\left({n j\pi\over L+1}\right) \\ 
& = & \sum_{k,j=1}^L {A_{kj} \over 1- {1 \over 2}
\left[ 
\cos\left({k \pi \over L+1} \right) +
\cos\left({j \pi \over L+1} \right) 
\right]}
\sin\left({m k\pi\over L+1}\right)
\sin\left({n j\pi\over L+1}\right) ~ .
\eea
Hence
\be
A_{kj}^S = N {A_{kj}^U \over
1-{1 \over 2}\left[ 
\cos\left({k \pi \over L+1} \right) +
\cos\left({j \pi \over L+1} \right) 
\right]}
\ee
The numerical prefactor $N$ can be determined by imposing that the
sum over $m$ and $n$ of $\ps_{m,n}$ is 1, that is
\be
\sum_{m,n} \sum_{k,j} A_{kj}^S \sin \left({m k \pi \over L+1} \right)
\sin \left({n j \pi \over L+1} \right) = 1 ~,
\ee
which implies
\be
{1 \over N} = \left[{L (L+1) \over 2} \right]^2 
\sum_{k,j} {(A_{kj}^U)^2 \over 1-{1 \over 2}\left[ 
\cos\left({k \pi \over L+1} \right) +
\cos\left({j \pi \over L+1} \right) 
\right]}~ .
\ee
In the limit of large $L$,
\be
\tres = {1 \over N} \simeq  \left[{L (L+1) \over 2} \right]^2
{(A_{11}^U)^2 \over 1-\cos\left({\pi \over L+1}\right)}
\simeq {2^7 \over \pi^6} L^2 ~.
\ee
Hence in the continuum $\tres \simeq \fra{32}{\pi^6}\fra{L^2}{D}$ and
\be
\beta \simeq  {32 \over \pi^6} ~ .
\label{A54}
\ee

In the limit of strong but finite barriers one finds
$\tres = L \les/(4D)$, so that $\alpha=1/4$.

\section{The effective distribution}
\label{App_eff}

We want to evaluate the effective distribution:
\be
\peffn \equiv \sum_{\tau=0}^\infty \Pdep(\tau) p_n(\tau) ~ ,
\label{eq:peff}
\ee
introduced in Sec.~\ref{riduzione}.
Since $p_n(\tau)$ decays to zero after a time of order $\tres$,
for regimes {\em i} and {\em ii} (where $\tres \ll \tdep$)
$\Pdep(\tau)$ can be taken as a constant. Hence
\be
\peffn = {1 \over \tdep} \sum_{\tau=0}^\infty p_n(\tau) ~ .
\hbox{~~~~~~({\em i} and {\em ii})}
\ee

The sum $\sum_{\tau} p_n(\tau)$ has been shown in App.~\ref{App_rho}
to be equal to the solution $\rho_n$ of the stationary diffusion equation,
which always has a parabolic shape. Its normalized version 
$\ps_n$ is
\be
\ps_n = {\rho_n \over \sum_{n=1}^L \rho_n} = {\rho_n \over \tres} ~ ,
\ee
so that
\be
\peffn = {\tres \over \tdep} \ps_n ~ .
\hbox{~~~~~~({\em i} and {\em ii})}
\ee

This equation corresponds to Eq.~(\ref{hatp2}) in the limit $\tres \ll \tdep$.

For strong and infinite barriers (regimes {\it ii} and {\em iii}),
the contribution of times shorter than $\ttr$ is smaller than
$\ttr/\tdep$ and therefore negligible.
Accordingly, we can evaluate the sum (\ref{eq:peff}) using the
expression for $p_n(\tau)$ that is valid in the limit $\tau\gg\ttr$
[see Eq.~(\ref{eq:pnt}), in $d=2$ the generalization is trivial], 
$p_n(\tau)=(1/L^d)\exp(-\tau/\tres)$ and obtain
\be
\peffn = {1 \over L^d \tdep} \sum_{\tau=0}^\infty
\exp\left[-\tau \left({1 \over \tdep}+ {1 \over \tres} \right)
\right] ~ .
\ee
Converting the sum over discrete times into an integral we have
\be
\peffn = {\tres \over \tres + \tdep} {1 \over L^d} =
{\tres \over \tres + \tdep} \ps_n ~ ,
\label{genericahatp}
\ee
where -- as usual -- $\ps_n$ is the normalized solution
of the stationary diffusion equation.
Thus, formula (\ref{hatp2}):
\be
\peffn = {\tres \over \tres + \tdep} \ps_n 
\nonumber
\ee
can be used in all the different regimes.

In the continuum it is possible to work out a more rigorous approach
and determine $\peffn$ as the
solution of a differential equation, which is the
generalization of the stationary diffusion equation (\ref{eq_dif}).
We start with the diffusion equation for $p(x,t)$, 
$\partial_t p=D\nabla^2 p$, where $x$ is a
$d-$dimensional vector. If we 
multiply both sides by $\Pdep(t)$ and integrate in time, we obtain:
\be
\int_0^\infty dt \Pdep(t) \partial_t p(x,t) = D\nabla^2
\int_0^\infty dt \Pdep(t) p(x,t) ~ .
\label{edi}
\ee

The right hand side is just $D\nabla^2 \peffx$ while the left
hand side is:
\bea
&& \int_0^\infty dt \Pdep(t) \partial_t p(x,t) \\
&=& \left. \Pdep(t) p(x,t)\right|_0^\infty -
   \int_0^\infty dt [\partial_t \Pdep(t)] p(x,t) \\
&=& -{1\over \tdep}{1\over {\cal A}} + {1\over \tdep} 
\int_0^\infty dt \Pdep(t) p(x,t) 
\eea
and Eq.~(\ref{edi}) becomes:
\be
D\nabla^2 \peffx - {1\over \tau_{dep}} \peffx + F = 0 ~ .
\label{eqgen}
\ee

It differs from the stationary diffusion equation $D\nabla^2\rho + F=0$
because of the presence of a `desorption' term ($-\peffx/\tau_{dep}$)
which is the responsible of the saturation of $\peffx$ at large $\les$.
As a matter of fact, in the limit $\les\to\infty$, $\rho$ is
known to diverge as $\fra{F}{D}L\les$ [see 
Eqs.~(\ref{rho_1d},\ref{rho_2d})] while the above equation
clearly shows that $\peffx$ goes to the constant 
$F\tdep=1/{\cal A}$.

The exact solution of (\ref{eqgen}) can be found both in $d=1$ and in
$d=2$ for a circular terrace and the proof that
$\peffx = \fra{\tres}{\tres+\tdep}\ps(x)$ works much in the
same way in the two cases. We give here a few more details for the
bidimensional case.
The solution of Eq.~(\ref{eqgen}) with the usual boundary condition
$\partial_r\peff(r)=-\peff(r)/\les$ evaluated for $r=L$ (the radius
of the circular terrace) is
\be
\peff(r) = {1\over \pi L^2} \left[
1 - { I_0(\fra{r}{\sqrt{D\tdep}}) \over
      I_0(\fra{L}{\sqrt{D\tdep}}) + \fra{\les}{\sqrt{D\tdep}}
      I_1(\fra{L}{\sqrt{D\tdep}}) }   \right] ~ ,
\ee
where $I_0$ and $I_1$ are the modified Bessel functions of order zero and one,
respectively. The arguments of the Bessel functions are at most
equal to $(\fra{L}{\sqrt{D\tdep}})=\sqrt{8\ttr\over\tdep}$, a small
quantity. An expansion of the Bessel functions gives:
\be
\peff(r) = {1\over \pi L^2} \,
           { L^2 + 2\les L - r^2 \over
             4D\tdep + L(L+2\les)  } ~ .
\ee

By using the results (\ref{cont_2d}) and (\ref{rho_2d}), after some algebra
we obtain the final expression:
\be
\peff(r) = {\tres \over \tdep + \fra{L}{D}(\fra{L}{4}+\fra{\les}{2}) }
\ps(r) ~ , ~~~d=2 ~ .
\ee

The calculation in $d=1$ leads to the result:
\be
\peff(x) = {\tres \over \tdep + \fra{L}{D}(\fra{L}{8}+\fra{\les}{2}) }
\ps(x) ~ , ~~~d=1 ~ .
\ee

The quantity $\fra{L}{D}(\cdots)$ appearing on the right hand side
at denominator does not coincide
with $\tres$ because the term $\fra{L^2}{D}$ has a prefactor
$\fra{1}{4}$ instead that $\fra{1}{8}$ in $d=2$ and a prefactor
$\fra{1}{8}$ instead that $\fra{1}{12}$ in $d=1$.
Nonetheless such quantity differs from $\tres$ for a quantity of order
$\ttr$ which can be safely neglected with respect to $\tdep$
(always appearing at denominator) so that, in the limit
$\ttr\ll\tdep$ (a limit applied throughout the paper)
we can conclude that the relation
\be
\peffx = {\tres \over \tdep + \tres} \ps(x) 
\ee
is always valid.


\begin{references}
\bibitem{nota_sigma}
It is easy to realize that $\omf$ should read $\omf=\fra{1}{\Delta t}
\bar\rho^2$, where $\Delta t=1/2dD$ is the time an atom spends on
each lattice site ($d$ is the substrate dimension).
\bibitem{nota_dep}
In the discrete representation, if $p_0$ is the deposition
probability per time step we have $\Pdep(t)=p_0(1-p_0)^t$.
The definition of $\tdep$ as the average time of deposition implies
the relation $\tdep = (1-p_0)/p_0$, so that
$\Pdep(t)=(1+\tdep)^{-1}[(1+\tdep^{-1})^{\tdep}]^{-t/\tdep}$.
If $\tdep\gg 1$ (i. e. $\tdep\gg 1/(2dD)$ in the continuum picture)
the prefactor is $\tdep^{-1}$ and the quantity in square brackets is
just $e$. Hence $\Pdep(t)$ coincides with Eq.~(\ref{Pdep}).
\bibitem{nota_UU}
We take the single walker to be distributed uniformly at $t=0$
[$p_n(0)=\pu_n$], but a different choice changes $\Ndis$ and
$\Nall$ only slightly. For this reason the relations between
$W$ are $\Wni$ and $\Ndis$ and $\Nall$ are given by the
symbol $\simeq$.
\end{references}
\end{document}